\documentclass[11pt,paper]{JHEP3}
\date{March 2008}
\usepackage{epsfig}
\usepackage{latexsym}
\usepackage{amssymb}
\usepackage{booktabs}
\newcommand{\be}{\begin{equation}}
\newcommand{\ee}{\end{equation}}
\newcommand{\ba}{\begin{eqnarray}}
\newcommand{\ea}{\end{eqnarray}}
\newcommand{\bi}{\begin{itemize}}
\newcommand{\ei}{\end{itemize}}
\newcommand{\tr}{{\rm Tr\,}}
\newcommand{\re}{\mathop{\rm Re}}

\newcommand{\nn}{\nonumber \\}

\newcommand{\half}{{\textstyle\frac{1}{2}}}

\newcommand{\<}{\langle}
\renewcommand{\>}{\rangle}
\newcommand{\eq}{Eq.~}

\newcommand{\fig}{Fig.~}

\newcommand{\la}{\label}

\newcommand{\txts}{\textstyle}

\newcommand{\hmu}{\hat\mu}
\newcommand{\hnu}{\hat\nu}

\newcommand{\as}{a_{\sigma}}
\newcommand{\at}{a_{\tau}}
\newcommand{\betas}{\beta_{\sigma}}
\newcommand{\betat}{\beta_{\tau}}

\newcommand{\Ss}{S_{\sigma}}
\newcommand{\St}{S_{\tau}}

\newcommand{\Nt}{N_{\tau}}

\title{Cutoff effects on energy-momentum tensor correlators in lattice gauge theory}
\author{
Harvey~B.~Meyer\\
Center for Theoretical Physics\\
Massachusetts Institute of Technology\\
Cambridge, MA 02139, U.S.A.\\
E-mail: \email{meyerh@mit.edu}
}

\preprint{MIT-CTP 4029}
\abstract{
We investigate the  discretization errors affecting correlators of the 
energy-momentum tensor $T_{\mu\nu}$ at finite temperature in SU($N_c$) gauge theory
with the Wilson action and two different discretizations of $T_{\mu\nu}$.
We do so by using lattice perturbation theory and non-perturbative Monte-Carlo simulations.
These correlators, which are functions of Euclidean time $x_0$ and spatial momentum ${\bf p}$, 
are the starting point for a lattice study of the transport properties
of the gluon plasma. We find that the correlator of the energy $\int d^3x\, T_{00}$ 
has much larger discretization errors than the correlator of momentum  $\int d^3x\, T_{0k}$.
Secondly, the shear and diagonal stress correlators ($T_{12}$ and $T_{kk}$) require 
$\Nt\geq 8$ for the $Tx_0=\frac{1}{2}$ point to be in the scaling region and the cutoff 
effect to be less than $10\%$. We then show that their discretization errors 
on an anisotropic lattice with $\as/\at=2$ are comparable to those on the isotropic lattice
with the same temporal lattice spacing. Finally, we also study finite ${\bf p}$ correlators.
}

\begin{document}
\section{Introduction}
Energy-momentum tensor (EMT) correlators in finite-temperature 
QCD provide a way to study  the spatial correlation of fluctuations in energy density, 
pressure and entropy~\cite{Meyer:2008dt}. 
They thus allow us to gain insight into the structure of 
the quark-gluon plasma (QGP), thereby going beyond its thermodynamic properties.
The time-dependent EMT correlators are related to the transport properties of the 
QGP. Phenomenological upper bounds for the shear
viscosity to entropy density ratio, $\eta/s<\frac{5}{4\pi}$
(see~\cite{Song:2008hj} and Refs. therein), are derived by comparing
hydrodynamic calculations of elliptic flow to heavy ion collision data.
This result suggests the picture of a strongly coupled plasma around $2T_c$,
where $T_c$ is the QCD crossover temperature.
A significant effort is underway to constrain 
the transport properties of the gluonic sector 
non-perturbatively from first principles~\cite{Nakamura:2004sy,Meyer:2008sn},
and given the  high computational cost of 
determining the EMT correlators in Monte-Carlo simulations,
it is important to optimize the choice of lattice action
and discretization of the EMT. This is the subject of this paper.

The gluonic correlators have been computed at treelevel 
in the continuum~\cite{Meyer:2008gt}.
Here we compute the correlators at finite lattice spacing 
using the (anisotropic) Wilson plaquette action~\cite{Wilson:1974sk} 
and two discretizations of the EMT.
The strategy is to determine the kinematic regime where
the treelevel discretization errors are below the $10\%$ level.
We then expect treelevel improvement to further reduce the 
discretization errors to the few percent level.
This is a reasonable target, given the statistical accuracy
being currently achieved in lattice simulations~\cite{Meyer:2008sn}.

The outline of this paper is as follows. In section 2 we give the definitions
of the lattice actions and discretizations of the EMT, and derive the treelevel formulas
for the two-point functions on the lattice. Section 3 is devoted to analyzing 
the treelevel cutoff effects for various correlators. Section 4 describes how the treelevel 
discretization errors can be removed from non-perturbative Monte-Carlo data, 
and in section 5 we discuss to what extent treelevel improvement is successful.
We end with some concluding remarks.

\section{Definitions and master formulas}

\subsection{In the continuum}
The continuum Euclidean energy-momentum tensor for SU($N_c$) gauge theories reads
\ba
T_{\mu\nu}(x) &=& \theta_{\mu\nu}(x) + \frac{1}{4}\delta_{\mu\nu}\,\theta(x)
\\
\theta_{\mu\nu}(x) &=&
 {\txts\frac{1}{4}}\delta_{\mu\nu}F_{\rho\sigma}^a F_{\rho\sigma}^a
   - F_{\mu\alpha}^a F_{\nu\alpha}^a
\\
\theta(x) &=& {\txts\frac{\beta(g)}{2g}} ~ F_{\rho\sigma}^a  F_{\rho\sigma}^a
\\
\beta(g) &=&  -b_0g^3+\dots,\qquad  b_0={\txts\frac{11N_c}{3(4\pi)^{2}}}\,.
\ea
We will study the dimensionless, finite-temperature $T$ Euclidean correlators
\be
C_{\mu\nu,\rho\sigma}(x_0,{\bf p}) \equiv T^{-5} 
\int d^3{\bf x}\,\, e^{i{\bf p\cdot x}}\,\<T_{\mu\nu}(x_0,{\bf x}) \,T_{\rho\sigma}(0)\>\,.
\la{eq:def}
\ee
$C_{\mu\nu,\mu\nu}$ (no summation) will be sometimes abbreviated as $C_{\mu\nu}$,
and $C_{\theta\theta}$ is the correlator of the trace anomaly $\theta$, normalized as in 
\eq(\ref{eq:def}).
In the continuum, these correlators were calculated to leading order in perturbation theory 
in~\cite{Meyer:2008gt}. The extent of the time direction is denoted by $L_0\equiv 1/T$.

\subsection{On the lattice}
On the anisotropic lattice with spatial lattice spacing $\as$ and temporal lattice spacing $\at$,
the Wilson action reads
\be
S_{\rm g} =   \sum_x \betas \Ss(x)  + \betat \St(x).
\ee
In numerical practice, it is convenient to parametrize these parameters as
\be
\betas=\frac{\beta}{\xi_0} \qquad\qquad  \betat=\beta\xi_0.
\ee
At treelevel, $\xi_0=\as/\at$. We use the notation
\ba
\Ss= \sum_{k<l} S_{kl},& ~ &\St =   \sum_{k} S_{0k}\,,\qquad
S_{\mu\nu}(x) = \textstyle{\frac{1}{N_c}} \re\tr\{1-P_{\mu\nu}(x)\}, \\
P_{\mu\nu}(x)&=& U_{\mu}(x)U_\nu(x+a_\mu\hat\mu)U_\mu(x+a_\nu\hat\nu)^{-1}U_\nu(x)^{-1}\,.
\ea

% First we introduce our notation. 
The lattice spacing in the four directions
are denoted by $a_\mu$ in order to maintain the symmetry among different directions 
as long as possible. At the end of the calculation we will set $a_1=a_2=a_3=\as$ and $a_0=\at$.
The vectors $\hmu$ are defined as unit vectors along the lattice axes.
We employ the summation convention for color indices, but not for space-time 
indices. We use the standard notations
\ba
\partial_\mu f(x) &=& \frac{1}{a_\mu}(f(x+a_\mu\hat\mu)-f(x)),\quad
\partial^*_\mu f(x) = \frac{1}{a_\mu}(f(x)-f(x-a_\mu\hat\mu)),\quad
\widetilde\partial_\mu = \half(\partial_\mu+\partial^*_\mu )\,,\qquad\\
\hat p_\mu &=& \frac{2}{a_\mu} \sin \frac{a_\mu p_\mu}{2},\qquad
\dot p_\mu = \frac{1}{a_\mu} \sin a_\mu p_\mu,\qquad
\hat p^2 = \sum_{\mu=0}^3 \hat p_\mu ^2\,.
\ea
% The vectors $\hmu$ are defined as unit vectors along the lattice axes.
%  with length $a_\mu$.

We introduce the perturbative fields $A_\mu(x)$ by
\be
U_\mu(x)=e^{g_0a_\mu A_\mu(x)}.
\ee
Using an antihermitian set of generators normalized by $\tr\{T^aT^b\}=-\frac{\delta_{ab}}{2}$,
we define $A_\mu(x)=A_\mu^a(x)T^a$. The latter has the covariant-gauge propagator
\be
\<\,A_\mu^a(x)\,A_\nu^b(y)\,\>_0 = \delta^{ab}
\int_{\cal B} \frac{d^4p}{(2\pi)^4}
\frac{e^{i(p(x-y)+\frac{1}{2}a_\mu p_\mu - \frac{1}{2} a_\nu p_\nu)}}{\hat p^2} ~ 
\left\{\delta_{\mu\nu} - (1-\lambda_0^{-1}) \frac{\hat p_\mu \hat p_\nu}{\hat p^2}\right\}.
\ee
The expectation value $\<\dots\>_0$ is taken in the free theory.
Here and in the following,
$\<O_1 O_2\>$ (whether in the free theory or not)
will always be understood to be the connected two-point function.
The Brillouin zone is ${\cal B}={\otimes}_{\mu} {\cal B}_\mu$
with ${\cal B}_\mu = [-\pi/a_\mu,\pi/a_\mu]$.
The lattice field strength is defined by
\be
P_{\mu\nu}(x) % = U_\mu(x) U_\nu(x+\hat\mu) U_\mu(x+\hat\nu)^{-1}U_\nu(x)^{-1}
\equiv \exp\{g_0a_\mu a_\nu F_{\mu\nu}(x)\}.
\ee
To leading order, one has 
\be
F_{\mu\nu}(x) = \partial_\mu A_\nu(x)-\partial_\nu A_\mu(x)+ {\rm O}(g_0).
\ee
%%%%%%%%%%%%%%%%%%%%%%%%%%%%%%%%%%%%%%%%%%%%%%
\subsection{Clover discretization}
%%%%%%%%%%%%%%%%%%%%%%%%%%%%%%%%%%%%%%%%%%%%%%
We  introduce $Q_{\mu\nu}(x)$ and $\widehat F_{\mu\nu}(x)$ as in~\cite{Luscher:1996sc}, 
\ba
Q_{\mu\nu}(x) &=& P_{\mu\nu}(x)+P_{\mu\nu}(x-a_\mu\hmu)+
 P_{\mu\nu}(x-a_\nu\hnu)+ P_{\mu\nu}(x-a_\mu\hmu-a_\nu\hnu),\\
\widehat F_{\mu\nu}(x) &=&  \frac{1}{8a_\mu a_\nu}(Q_{\mu\nu}(x)-Q_{\nu\mu}(x)).
\ea
We define $\widehat F_{\mu\nu}= g_0\widehat F_{\mu\nu}^a T^a$, implying
for instance
$-\frac{2}{g_0^2}\tr\{\widehat F_{\mu\nu}(x)\widehat F_{\mu\nu}(x)\}= 
\widehat  F_{\mu\nu}^a(x)\widehat   F_{\mu\nu}^a(x)$.
Then 
\ba
\widehat F^a_{\mu\nu}(x) &=& \frac{1}{2}\left[
\widetilde\partial_\mu(A^a_\nu(x)+A^a_\nu(x-\hat\nu))-
\widetilde\partial_\nu(A^a_\mu(x)+A^a_\mu(x-\hat\mu))
 +{\rm O}(g_0) \right].
\nonumber
\ea

Using Wick's theorem, one  finds\footnote{If $O_{i}^a$ are linear combinations of the gauge fields,
$O_i^a=\lambda_{i,\alpha}A_\alpha^a$,
then by Wick's theorem
\ba
&&\sum_{a,b}\<(O_1^a O_2^a) ~ (O_3^b O_4^b) \>_{\rm 0,conn} =  
\sum_a \<O_1^a O_3^a\>_0 \<O_2^a O_4^a\>_0 +  \<O_1^a O_4^a\>_0 \<O_2^a O_3^a\>_0 \nn
&& = d_A \,(\, \< O_1 O_3\> \< O_2 O_4\> \,+\,  \< O_1 O_4\>  \< O_2 O_3\>\,)_{\rm U(1)} .
\nonumber
\ea
The second equality follows from the fact that $\<O_1^a O_2^a\>_0$
is independent of $a$ and equal to the corresponding correlation function in the U(1) gauge theory.}
($d_A=N_c^2-1$)
\ba
&& \<~ \widehat F^a_{\mu\nu}(x)\widehat F^a_{\rho\sigma}(x)~
   \widehat F^b_{\alpha\beta}(y)\widehat F^b_{\gamma\delta}(y) ~\>_0 \nn
 && = d_A\Big[ \< \widehat F_{\mu\nu}(x) \widehat F_{\alpha\beta}(y)\>_0
 \< \widehat F_{\rho\sigma}(x) \widehat F_{\gamma\delta}(y)\>_0
 + \< \widehat F_{\mu\nu}(x) \widehat F_{\gamma\delta}(y)\>_0
 \< \widehat F_{\rho\sigma}(x) \widehat F_{\alpha\beta}(y)\>_0 \Big|_{\rm U(1)}.
 \nonumber
\ea
Then, in Feynman gauge $\lambda_0=1$, we obtain
\ba
\< \widehat F_{\mu\nu}(x) \widehat F_{\alpha\beta}(y)\>
&=& d_A\,\phi_{\mu\nu\alpha\beta}(x-y)\,, \\
 \phi_{\mu\nu\alpha\beta}(x) &\equiv & 
   \delta_{\nu\beta}  f^\nu_{\mu\alpha}(x)
+ ~\delta_{\mu\alpha} f^\mu_{\nu\beta}(x)
- ~\delta_{\mu\beta} f^\mu_{\nu\alpha}(x)
- ~\delta_{\nu\alpha} f^\nu_{\mu\beta}(x)\,, \\
f^\nu_{\mu\alpha}(x)&\equiv & 
\int_{\cal B} \frac{d^4p}{(2\pi)^4} \frac{e^{ipx}}{\hat p^2}
~ \cos^2(p_\nu a_\nu/2) ~ \dot p_\alpha ~\dot p_\mu\,. 
\la{eq:f...}
\ea
Finally,
\be
\< \widehat F^a_{\mu\nu}(x)\widehat F^a_{\rho\sigma}(x)~
   \widehat F^b_{\alpha\beta}(y)\widehat F^b_{\gamma\delta}(y) \>_0 
 = d_A\Big[ \phi_{\mu\nu\alpha\beta}(x-y) \phi_{\rho\sigma\gamma\delta}(x-y)
  +   \phi_{\mu\nu\gamma\delta}(x-y)\phi_{\rho\sigma\alpha\beta}(x-y)\Big]\,.
\ee
This correlator is gauge-invariant and therefore independent of $\lambda_0$.
To go over to mixed propagators (which are functions of $(x_0,{\bf p})$),
we introduce the spatial Fourier transform of $\phi_{\mu\nu\alpha\beta}(x)$,
%\be
$\tilde \phi_{\mu\nu\alpha\beta}(x_0,{\bf p}) =
\as^3\sum_{\bf x} \phi_{\mu\nu\alpha\beta}(x) ~e^{i{\bf p\cdot x}}$.
%\ee
Then
\ba
\la{eq:masterII}
&& \as^3\sum_{\bf y} e^{i{\bf q}\cdot{\bf y}}~
\<~ \widehat F^a_{\mu\nu}(0)\widehat F^a_{\rho\sigma}(0)~ 
   \widehat F^b_{\alpha\beta}(x_0,{\bf y})\widehat F^b_{\gamma\delta}(x_0,{\bf y}) ~\>_0 \\
&& = d_A\int_{{\cal B}_\sigma} \frac{{\rm d}^3{\bf p}}{(2\pi)^3} 
\Big[\tilde\phi_{\mu\nu\alpha\beta}(x_0,{\bf p}) \tilde\phi_{\rho\sigma\gamma\delta}(x_0,-({\bf p+q}))
   + \tilde \phi_{\mu\nu\gamma\delta}(x_0,{\bf p})\tilde \phi_{\rho\sigma\alpha\beta}(x_0,-({\bf p+q}))\Big].
\nonumber
\ea
with  ${\cal B}_\sigma = {\cal B}_1\times {\cal B}_2\times {\cal B}_3$. 
\eq (\ref{eq:masterII}) is the master
formula from which we will derive all results in section 2.

Explicitly, in the clover case,
\ba
  \la{eq:masterI}
\tilde \phi_{\mu\nu\alpha\beta}(x_0,{\bf p}) 
 &=&  \int_{{\cal B}_0} \frac{dp_0}{2\pi} \frac{e^{ip_0x_0}}{\hat p_0^2+\hat{\bf p}^2}  \times \\
 & \times\Big[&\delta_{\nu\beta}\ \cos^2(a_\nu p_\nu/2) ~ \dot p_\mu ~ \dot p_\alpha 
      + ~\delta_{\mu\alpha} \cos^2(a_\mu p_\mu/2) ~ \dot p_\nu  ~ \dot p_\beta \nn
 &-&  \delta_{\mu\beta} ~ \cos^2(a_\mu p_\mu/2) ~ \dot p_\nu ~ \dot p_\alpha 
      - ~\delta_{\nu\alpha} \cos^2(a_\nu p_\nu/2) ~ \dot p_\mu ~ \dot p_\beta \Big]. 
 \nonumber
\ea
At finite temperature, one is to replace the $p_0$-integral in $\tilde\phi$ by a Matsubara sum,
\be
\int_{{\cal B}_0} \frac{dp_0}{2\pi} \to T\sum_{p_0\in{\cal B}_0}\,,
\qquad p_0=2\pi n T, \qquad n\in {\bf Z}.
\ee
It is worth noting that 
the scalar propagator in the mixed $(x_0,{\bf p})$ representation can be calculated 
explicitly, even at finite lattice spacing~\cite{Elze:1988zs}:
\be
\la{eq:elze}
\frac{1}{L_0}\sum_{p_0\in{\cal B}_0}
\frac{e^{ip_0x_0}}{\hat p_0^2 + \omega^2}
= \frac{\at}{2}\frac{\cosh(\hat\omega(\half L_0-x_0))}{\sinh(\hat\omega\at)\sinh(\half\hat\omega L_0)}\,,
\qquad
\frac{2}{a_\tau}\sinh\left(\frac{\hat\omega \at}{2}\right) = \omega\,.
\ee

\subsection{Plaquette discretization}
The gauge-invariant continuum operator $[F^a_{\mu\nu}(x)F^a_{\mu\nu}(x)]_{\rm cont}$ can be discretized 
gauge invariantly as 
\be
\frac{4}{g_0^2a_\mu^2 a_\nu^2}\re\tr\{1-P_{\mu\nu}(x)\} =
F^a_{\mu\nu}(x)F^a_{\mu\nu}(x) \, + \, {\rm O}(a_{\mu,\nu}^2)\,,
\la{eq:plaq}
\ee
by the lattice field $F^a_{\mu\nu}(x)$, where $F_{\mu\nu}(x)=F^a_{\mu\nu}(x)T^a$.
Then \eq(\ref{eq:masterII}) still applies if all the $\widehat F$ are replaced by $F$'s,
provided $f_{\mu\alpha}^\nu(x) $ (see \eq\ref{eq:f...}) is replaced by 
\be
f_{\mu\alpha}^\nu(x) = \int_{\cal B} \frac{d^4p}{(2\pi)^4} \frac{e^{ipx}}{\hat p^2}
~ e^{i(a_\alpha p_\alpha -p_\mu a_\mu)/2} ~ \dot p_\alpha ~\dot p_\mu\,. 
\ee

\section{Investigating cutoff effects}
In this section we investigate the numerical size of cutoff effects for two different 
discretization schemes of the energy momentum tensor correlators.
The length of the temporal direction is set equal to $L_0=1/T$, where $T$ is the temperature.
There is no finite temperature at which perturbation theory
correctly describes finite-volume effects, since those are related to 
magnetic screening, which is a non-perturbative effect. For that reason,
the spatial volume is kept infinite.

We use \eq(\ref{eq:elze}) and perform the spatial momentum integral 
in \eq(\ref{eq:masterII}) by Gaussian quadrature with a 
target relative accuracy of $10^{-4}$. 
We  checked a sample of the results against an extrapolation of 
the finite-volume momentum sums, and by comparing the latter sums to 
Monte-Carlo data at very high $\beta$ value.

\subsection{Correlators of conserved charges}
\FIGURE[t]{
\psfig{file=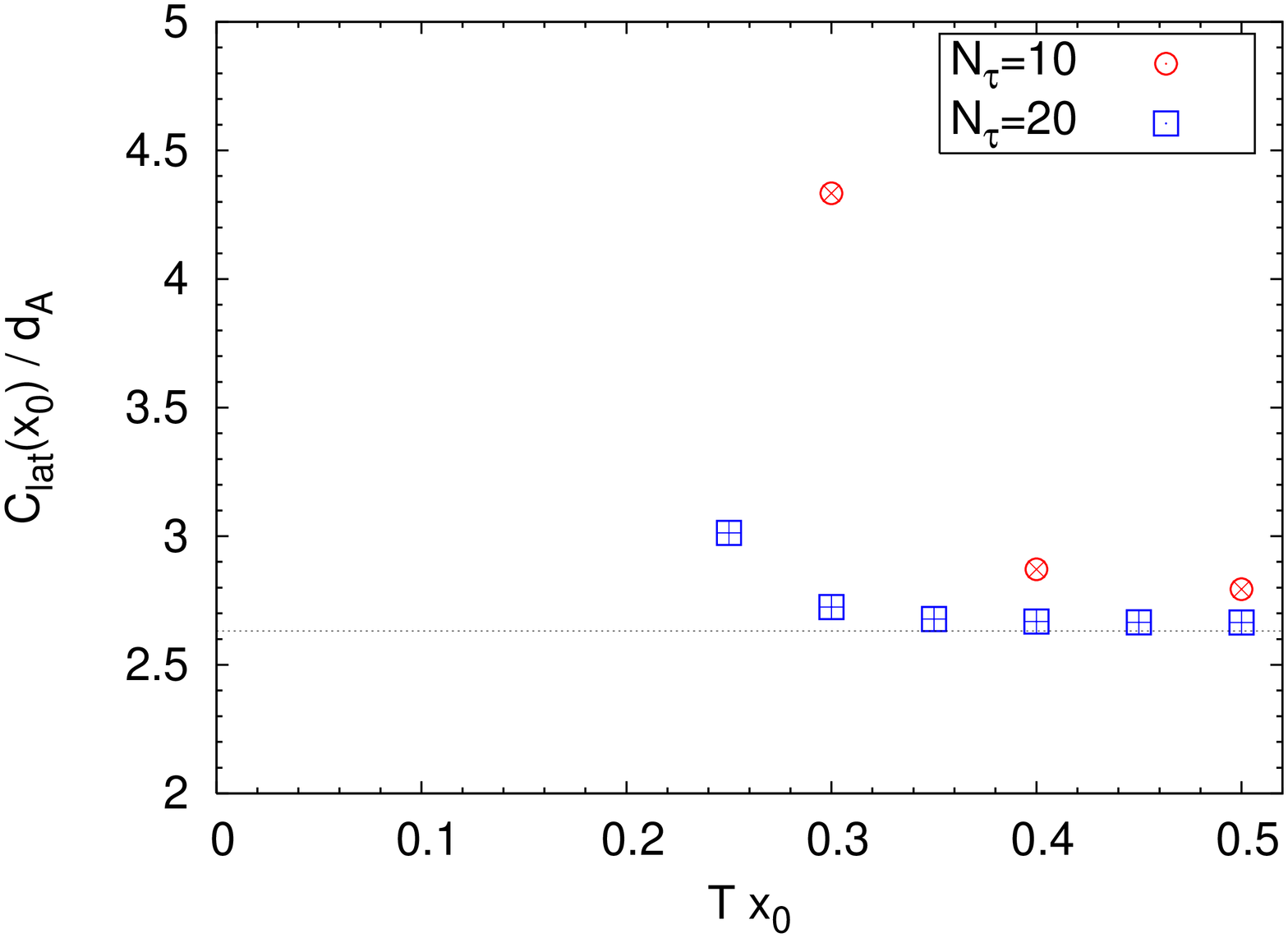,angle=-90,width=14.5cm}
\psfig{file=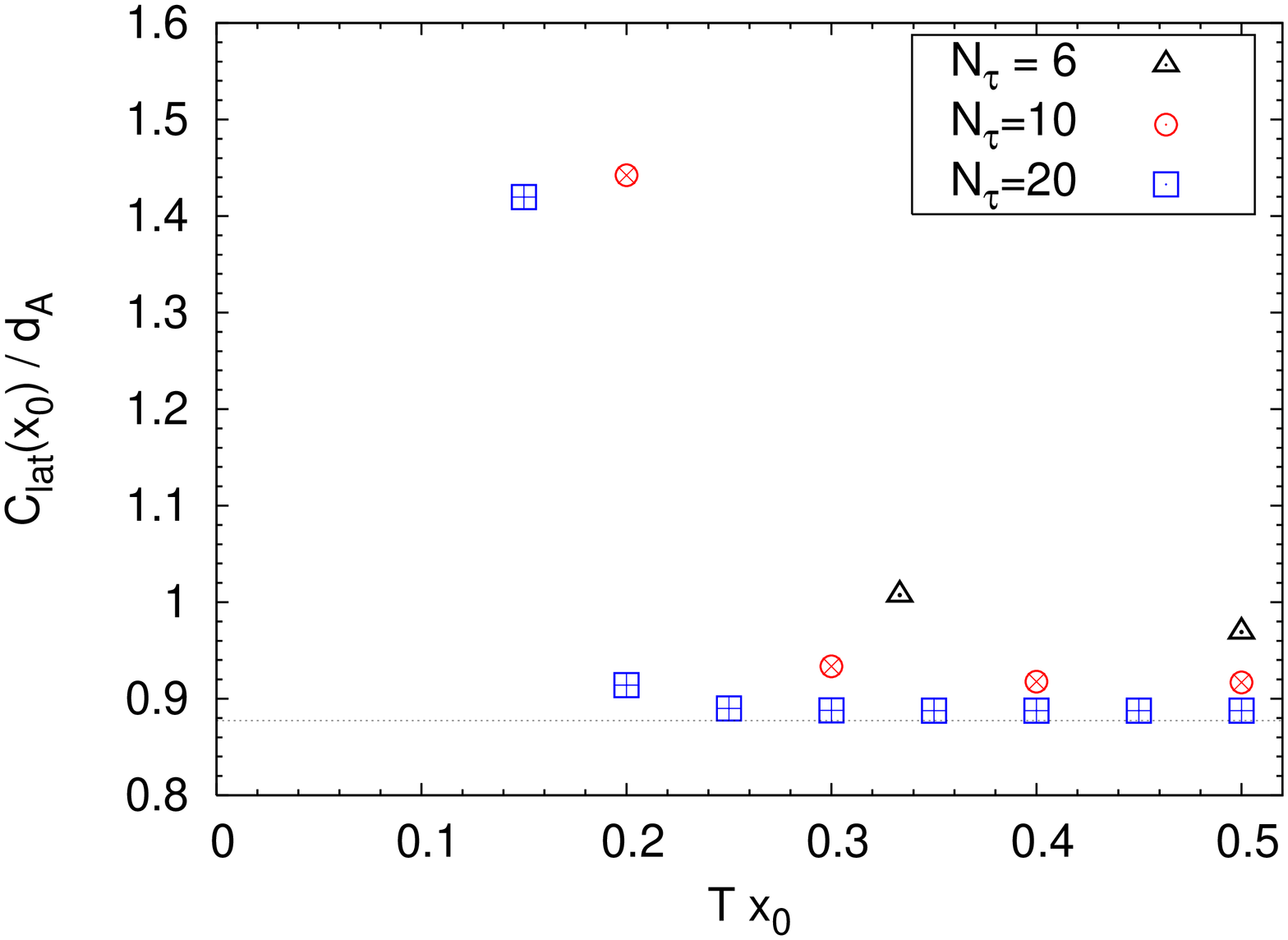,angle=-90,width=14.5cm}
\caption{{\bf Top}: the treelevel $C_{00,00}(x_0,{\bf 0})$ correlator
% $\as^3\sum_{\bf x} \<\theta_{00}(0)\,\theta_{00}(x)\>_0 $
correlator at finite lattice spacing on the isotropic lattice.
 The horizontal line is 
the continuum treelevel prediction, $4\pi^2/15$.
{\bf Bottom}: the treelevel $C_{03,03}(x_0,{\bf 0})$
% $\as^3\sum_{\bf x} \<T_{03}(0)\,T_{03}(x)\>_0 $
correlator at finite lattice spacing on the isotropic lattice.
 The horizontal line is 
the continuum trelevel prediction, $4\pi^2/45$.
%Notice the different scales on the vertical axes.
}
\la{fig:th00}
}
We start with the isotropic lattice.
Figure \ref{fig:th00} displays the treelevel lattice 
correlator of the energy operator, $\as^3\sum_{\bf x}T_{00}(x)$. 
The trace anomaly is formally O($\alpha_s$) and does not play a role 
at this leading order. Since in the continuum
\be
C_{00,00}(x_0,{\bf 0})=\frac{1}{T^5}\int d^3x\, \<T_{00}(x) T_{00}(0) \> =
\frac{c_v}{T^3}\qquad  \forall x_0\neq 0,
\ee
the departure of this correlator from a constant $c_v/T^3=4\pi^2d_A /15$ is a measure of 
discretization errors. The discretization errors fall below the $10\%$ level
  only for $x_0/a\geq 5$ for $\Nt=20$.
The reason for this large cutoff effect is that  generically
$C_{\mu\nu,\rho\sigma}(x_0,{\bf p})$  fall off as $x_0^{-5}$, and this singularity 
must cancel in the case of $\int d^3x T_{00}(x)$. It is not surprising that 
this large cancellation only takes place for $x_0\gg a$, where time-translation 
is effectively restored as a symmetry.

For the momentum density operator,
\be
\frac{1}{3}\sum_k C_{0k,0k}(x_0,{\bf 0})=\frac{1}{3T^5}\sum_k \int d^3x\, \<T_{0k}(x) T_{0k}(0) \> =
\frac{s}{T^3}\qquad  \forall x_0\neq 0,
\ee
the situation is significantly better (see \fig\ref{fig:th00}, bottom panel), 
although the same cancellation has to take place.
At $\Nt=20$, discretization errors are well below $10\%$ for $x_0\geq 4a$.

\subsection{The tensor channel ($\xi=1$)}
In the continuum and infinite spatial volume limit, the following equality holds
by rotational invariance:
\be
\frac{1}{4} \int d^3x\,\<( T_{11}- T_{22})(0)\,( T_{11}- T_{22})(x)\>
=\int d^3x\,\< T_{12}(0) T_{12}(x)\>\,,\quad \forall x_0.
\la{eq:T12id}
\ee
In finite spatial volume, the two correlators differ even in the continuum.
On the infinite cubic lattice, discretizing either side of \eq\ref{eq:T12id} yields
a correlator that approaches the continuum limit with different 
O($a^2$) discretization errors.
Figure~\ref{fig:T12conti} shows a comparison of the discretization errors
affecting these two schemes. With the clover discretization,
at a given value of $\Nt$, discretizing  $T_{12}$ 
yields smaller discretization errors than discretizing $\half(T_{11}-T_{22})$, 
for all values of $x_0$.

With the plaquette discretization, \eq\ref{eq:plaq}, 
one is only able to treat the diagonal elements of $T_{\mu\nu}$,
and in general it leads to significantly larger discretization errors than the clover
discretization. If $E$ refers to the chromo-electric field and $B$ to the magnetic field,
the $EE$ and $BB$ terms in the correlator are defined at integer values of $x_0/\at$,
while the $EB$ term is defined at half-integer values of $x_0/\at$.
% This is presumably because the magnetic part of the correlator
% is fudged over a time-interval of O($a$) according to $C_{BB}(x_0-a)+2C_{BB}(x_0)+C_{BB}(x_0+a)$.
For a function falling off as $x_0^{-5}$, without a careful treatment 
this mismatch leads to large O($a^2$) cutoff effects.
An appropriate scheme~\cite{Meyer:2007ic} is to compute separately $C_{BB}$, 
$C_{EE}$ at integer values of $x_0/a$ and $C_{EB}$ at half-integer values. 
One can then obtain the two-point function of $\half(T_{11}-T_{22})$ by interpolation, which is 
treated as part of the treelevel improvement scheme (to be discussed in section~\ref{sec:tlimp}). 
However, at treelevel, $C_{EB}$ vanishes at the midpoint $x_0=L_0/2$, and $C_{BB}=C_{EE}$, so it is 
straightforward to compare the plaquette discretization scheme to the others at that point. 
As figure~\ref{fig:T12conti} shows,
its cutoff effects  are almost identical to those obtained with the 
clover discretization of $T_{12}$.

In all cases, $\Nt\geq 8$ is necessary 
for the cutoff effects to be less than $10\%$ and in the O($a^2$) scaling region.

\FIGURE[t]{
\psfig{file=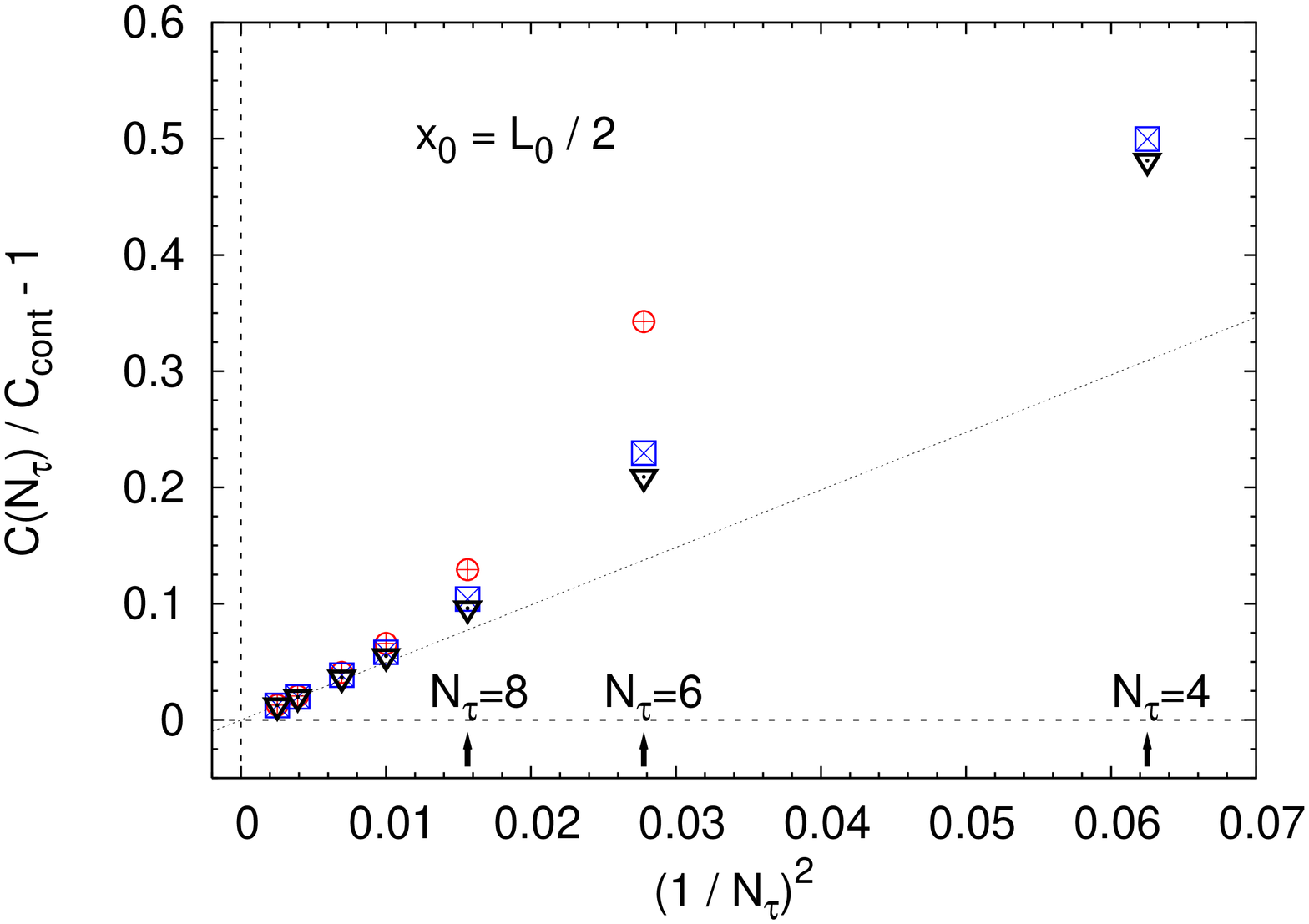,angle=-90,width=14cm}
\psfig{file=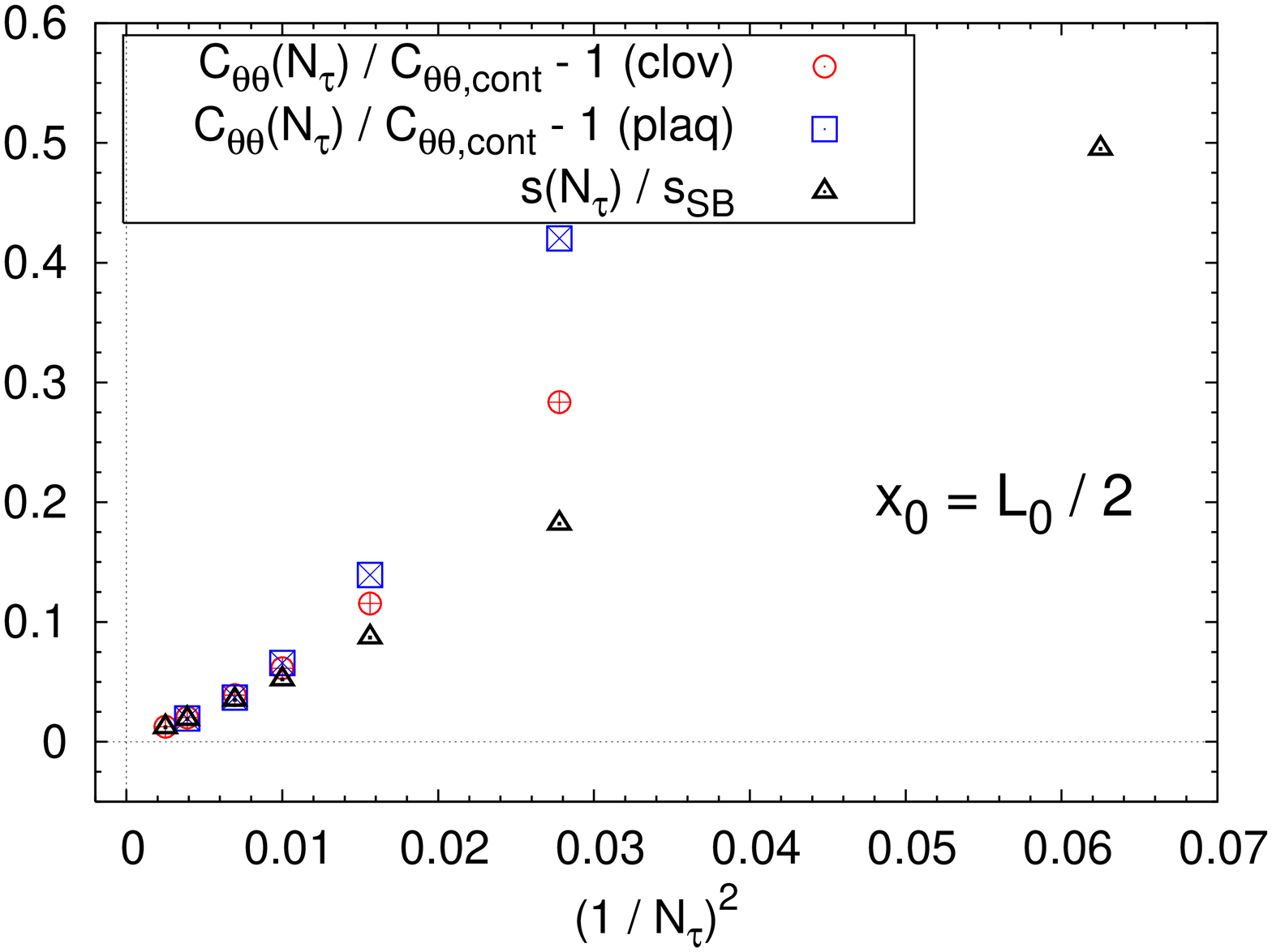,angle=-90,width=14cm}
\caption{{\bf Top}: the relative deviation of the ${\bf p}=0 $
tensor correlator at finite lattice spacing from the continuum correlator. 
The $\square$'s are for the clover discretization of $T_{12}$
and the $\otimes$ for the clover discretization of $\half(T_{11}-T_{22})$. 
The $\nabla$ is the  plaquette discretization.
{\bf Bottom}: the cutoff effects on the $\as^3\sum_{\bf x} \<\theta(0)\,\theta(x)\>_0 $
correlator  with the clover and plaquette discretizations.  
Note that $C_{\theta\theta}(\Nt)/C_{\theta\theta,{\rm cont}}  =  C_{00}(\Nt)/C_{00,{\rm cont}}$
at $x_0=L_0/2$. The entropy $s(\Nt)$ is computed with the standard Wilson action and plaquette
discretization of $\theta_{00}$~\cite{Beinlich:1995ik}.}
\la{fig:T12conti}
}

\subsection{The scalar channel $(\xi=1)$}
We consider two ways to evaluate the zero-momentum two-point function of $\sum_{k}T_{kk}$
on the lattice. They differ, for large $\Nt$ and $x_0/\at$, by ${\rm O}(a^2)$ terms.
\begin{enumerate}
\item The first way then consists in discretizing $\sum_{i,k} C_{ii,kk}$ directly.
 Using the notation introduced earlier, one may rewrite it identically as
 \be
\sum_{i,k} \int d^3x\, \<T_{ii}(0)\,T_{kk}(x)\>=
 \int d^3x \left( \<\theta_{00}(0)\,\theta_{00}(x)\>-\frac{3}{2}\<\theta(0)\,\theta_{00}(x)\>
           + \frac{9}{16} \<\theta(0)\,\theta(x)\>\right)\,.
 \la{eq:CiikkI}
 \ee
\item Alternatively one can exploit the conservation of energy to write
\be
\sum_{i,k} C_{ii,kk}(x_0,{\bf 0}) =
 \frac{6s-c_v}{T^3} + C_{\theta\theta}(x_0,{\bf 0})\,
\qquad (x_0>0).
\la{eq:TiiTmumu}
\ee
We have used the standard expressions for entropy and specific heat, 
$s=\frac{\partial p}{\partial T}$ and $c_v=\frac{\partial e}{\partial T}$.
Here one computes the trace-anomaly correlator in 
a given discretization, and subtracts the thermodynamic function appearing
on the right-hand side, either at the same value of $\Nt$ or 
already extrapolated to the continuum.
\end{enumerate}

\subsubsection{Asymptotic temperatures}
At high temperatures, $c_v \sim 3s$ and  we have
\be
\frac{6s-c_v}{T^3}= \frac{4\pi^2d_A}{15}
 \left[1-\frac{5N_c}{4}\frac{\alpha_s}{\pi}+{\rm O}(\alpha_s^{\frac{3}{2}}) \right].
\la{eq:thermo}
\ee
In particular, this quantity is positive and straightforward to compute non-perturbatively.
Since the $\theta$ two-point function is formally O($\alpha_s^2$),
\eq(\ref{eq:TiiTmumu}) implies that 
the leading expression for $C_{ii,kk}$ is simply \eq(\ref{eq:thermo}), independent
of $x_0$. The correlators in the scalar channel thus have a large $x_0$-independent contribution. 
At high temperatures, the choice between strategy (1.) and (2.) amounts to deciding
 which of $T^2(6s-c_v)$ or 
$\int d^3x \<\theta_{00}(0)\,\theta_{00}(x)\>_0$ has the smaller cutoff effects.
We expect the former to be the better quantity, since the latter correlator 
exhibits a contact term (as seen earlier), which  spreads over a fixed number of lattice spacings.
So provided the 
thermodynamic potentials are accurately known, the second strategy is the superior one.

It remains to be seen how large the cutoff effects on $\int d^3x \<\theta(0)\,\theta(x)\>_0$
are. This is shown on \fig\ref{fig:Sconti}. For $\Nt\geq 8$, they are comparable to the 
cutoff effects on the $T_{12}$ two-point function. Furthermore, they are not much larger
than the cutoff effects on the  entropy  computed with the standard
Wilson plaquette action~\cite{Beinlich:1995ik}, also displayed on the figure.

In summary, the trace-anomaly two-point function is  computationally advantageous in that 
one is computing directly a quantity which is already O($\alpha_s^2$). 

\subsubsection{Temperatures close to $T_c$}
Although perturbative methods fail near $T_c$, we know that very close to $T_c$,
the specific heat $c_v$ becomes large (both in SU(3) gauge theory and full QCD),
and  $6s-c_v$ is negative:
it cancels a large flat contribution in the trace-anomaly correlator~\cite{Huebner:2008as}. 
In that regime, it is therefore preferable to adopt the first strategy and 
compute $\sum_{i,k}C_{ii,kk}$ directly.

\subsection{Anisotropic lattice}
We now turn to the case of an anisotropic lattice, $\xi>1$.
Indeed it has long been recognized that such a lattice presents certain advantages 
for the calculation of thermodynamics~\cite{Namekawa:2001ih} and thermal correlation functions, 
in particular in charmonium calculations~\cite{Asakawa:2003re}.

In order to find the optimal range of anisotropies, we consider 
the cutoff effects on the tensor correlators at a fixed value of the 
spatial lattice spacing, $\as/L_0=$fixed. We then vary the anisotropy 
between 1 and 4. On \fig\ref{fig:T12aniso}, we see that the 
sign of the cutoff effects changes for $1<\xi<2$, and goes to a finite value 
in the Hamiltonian limit, $\xi\to\infty$. It is clearly seen that any choice $\xi\geq2$
reduces the cutoff effects significantly. It also appears that choosing $\xi>3$ does 
not reduce the cutoff effects further, presumably 
because they are dominated by the coarseness
of the spatial discretization. The cutoff effects appear to be minimal near $\xi=2$,
and we make the choice to investigate $\xi=2$ in the following.
In fact, as \fig(\ref{fig:T12aniso}) shows, 
$\Nt=16$ is about as good on the $\xi=2$ lattice as on the isotropic lattice.
Because of the sign change of the cutoff effect, the smallness of the cutoff effect 
may be partly accidental. 
However, we find that other correlators are also improved.
For instance, the discretization errors on the energy correlator,
which are large on the isotropic lattice (\fig\ref{fig:th00} bottom panel),
are significantly (\fig\ref{fig:T12aniso}) reduced at $\xi=2$.
For $\Nt=20$ on the anisotropic lattice, the discretization error
is below $10\%$ for $Tx_0\geq \frac{3}{10}$, which is not the case on the 
$\Nt=10$ isotropic lattice.
We remark that even in the large $\xi$ limit, the energy correlator 
with the present discretization is not flat, in spite of continuous 
time-translation invariance being restored in that limit.

\FIGURE[t]{
\psfig{file=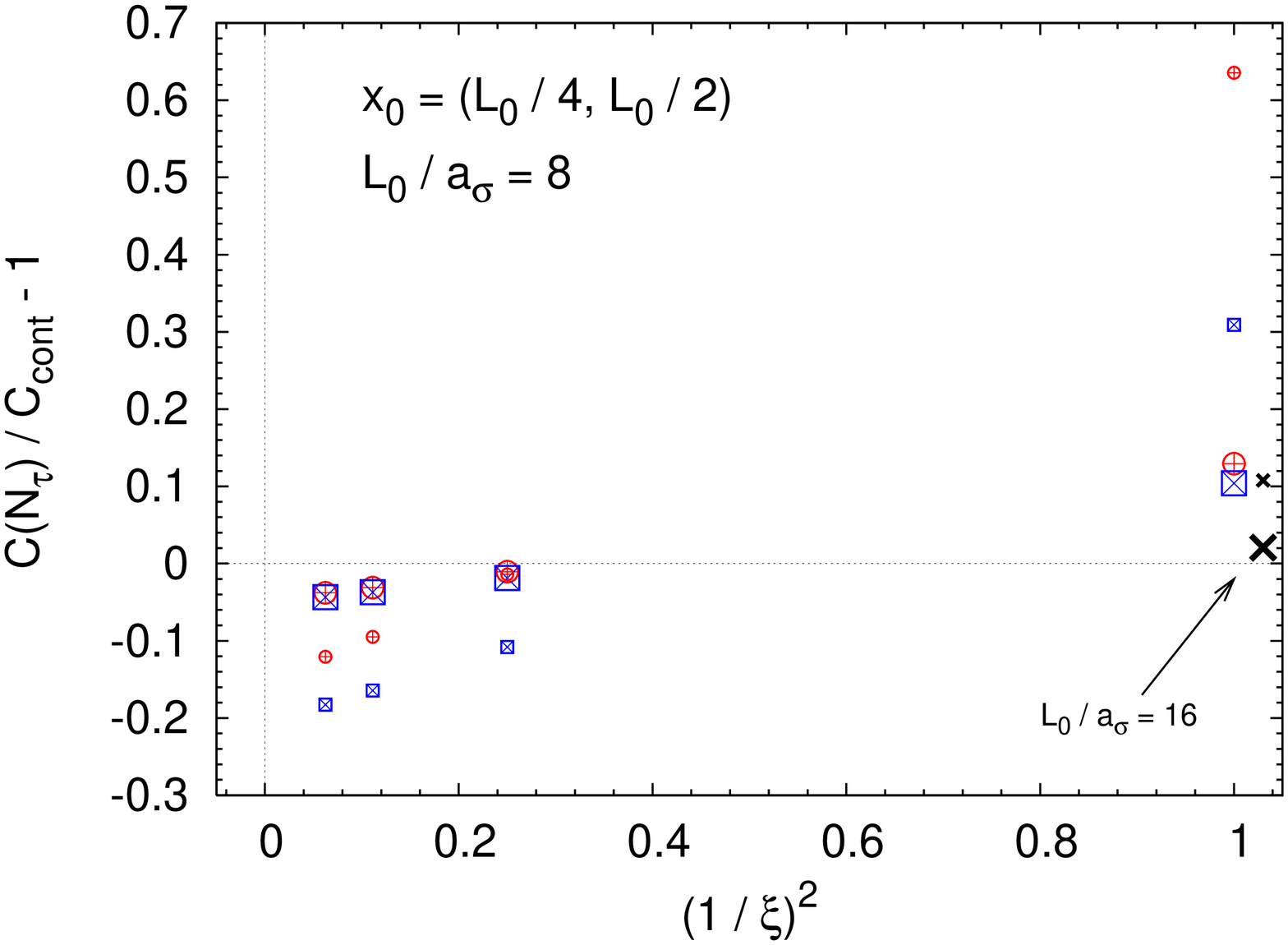,angle=-90,width=13cm}
\psfig{file=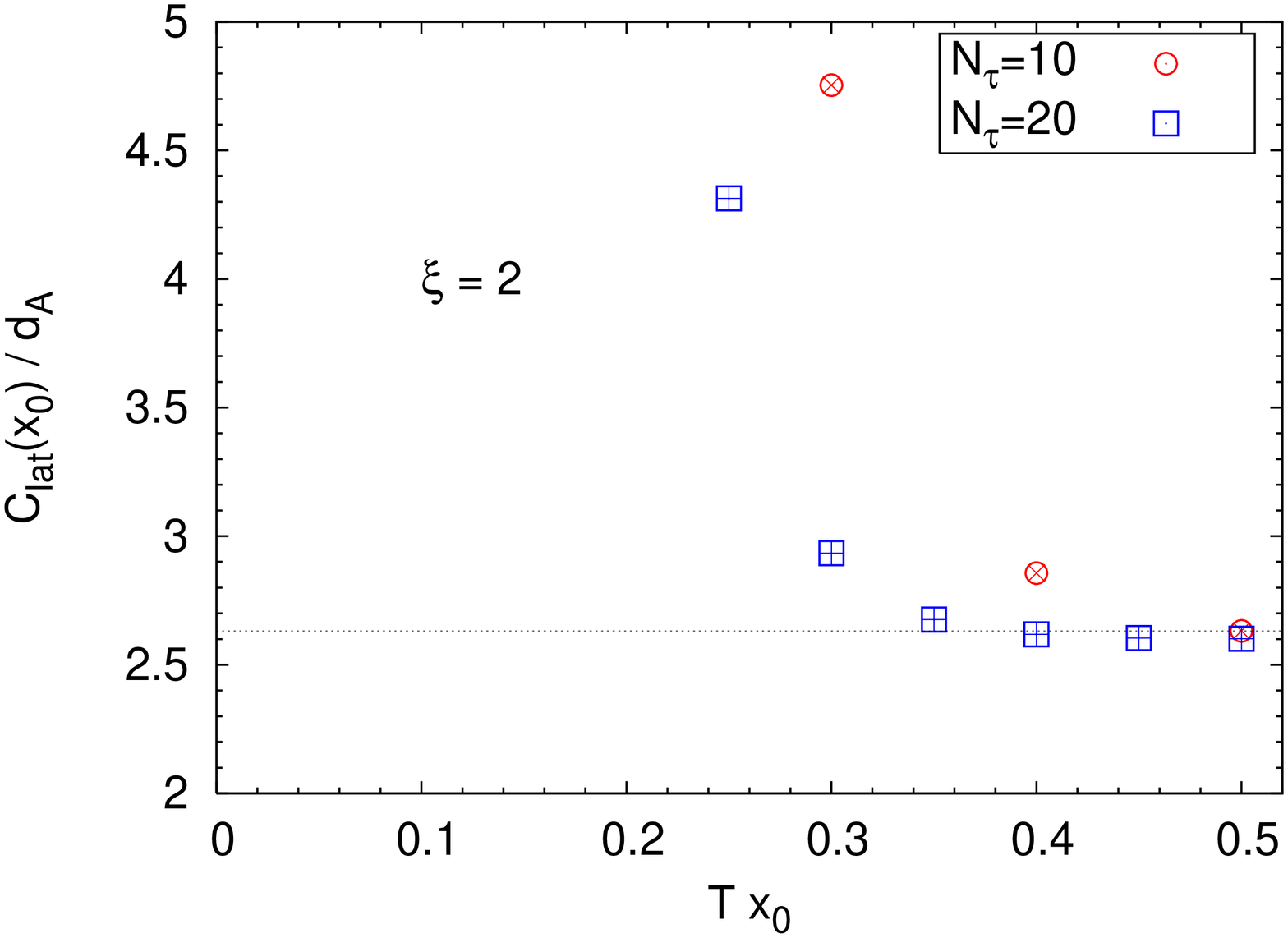,angle=-90,width=13cm}
\caption{{\bf Top}:  The cutoff effects on the tensor correlator for $x_0=L_0/4$ and $L_0/2$, 
corresponding to small and large symbols respectively. The spatial lattice spacing $\as$ is held fixed, 
the temporal lattice spacing $\at$ is varied between a quarter and one times $\as$.
The $\square$'s refer to $T_{12}$ and the $\otimes$'s to $\half(T_{11}-T_{22})$.
In addition, the two crosses at $\xi=1$ indicate the reduction of the cutoff effect 
on the $\half(T_{11}-T_{22})$ correlators when increasing $L_0/\as$ from 8 to 16.
{\bf Bottom}:  the treelevel $C_{00,00}(x_0,{\bf 0})$
correlator on the anisotropic lattice, to be compared with \fig(\ref{fig:th00}).
}
\la{fig:T12aniso}
}

\subsection{Non-zero spatial momentum}

For low momenta and frequencies, hydrodynamics predicts the functional 
form of the spectral functions in the shear channel $(\rho_{13,13})$ 
and the sound channel ($\rho_{33,33}$) (see e.g. \cite{Teaney:2006nc}).
It is therefore of interest to study also correlators with non-vanishing 
spatial momentum~\cite{Meyer:2008dq}. An example is shown on \fig\ref{fig:momentum}
for ${\bf p}=(0,0,\pi T)$.
Here too the cutoff effects are smaller for the momentum density
correlator than for the energy density correlator. 
For instance, at $\Nt=16$, the cutoff effects are less than $5\%$ for 
$Tx_0\geq \frac{1}{4}$ in the former case, 
while this level of accuracy only occurs for $Tx_0\geq \frac{3}{8}$ in the latter case.

\FIGURE[t]{
\centerline{\psfig{file=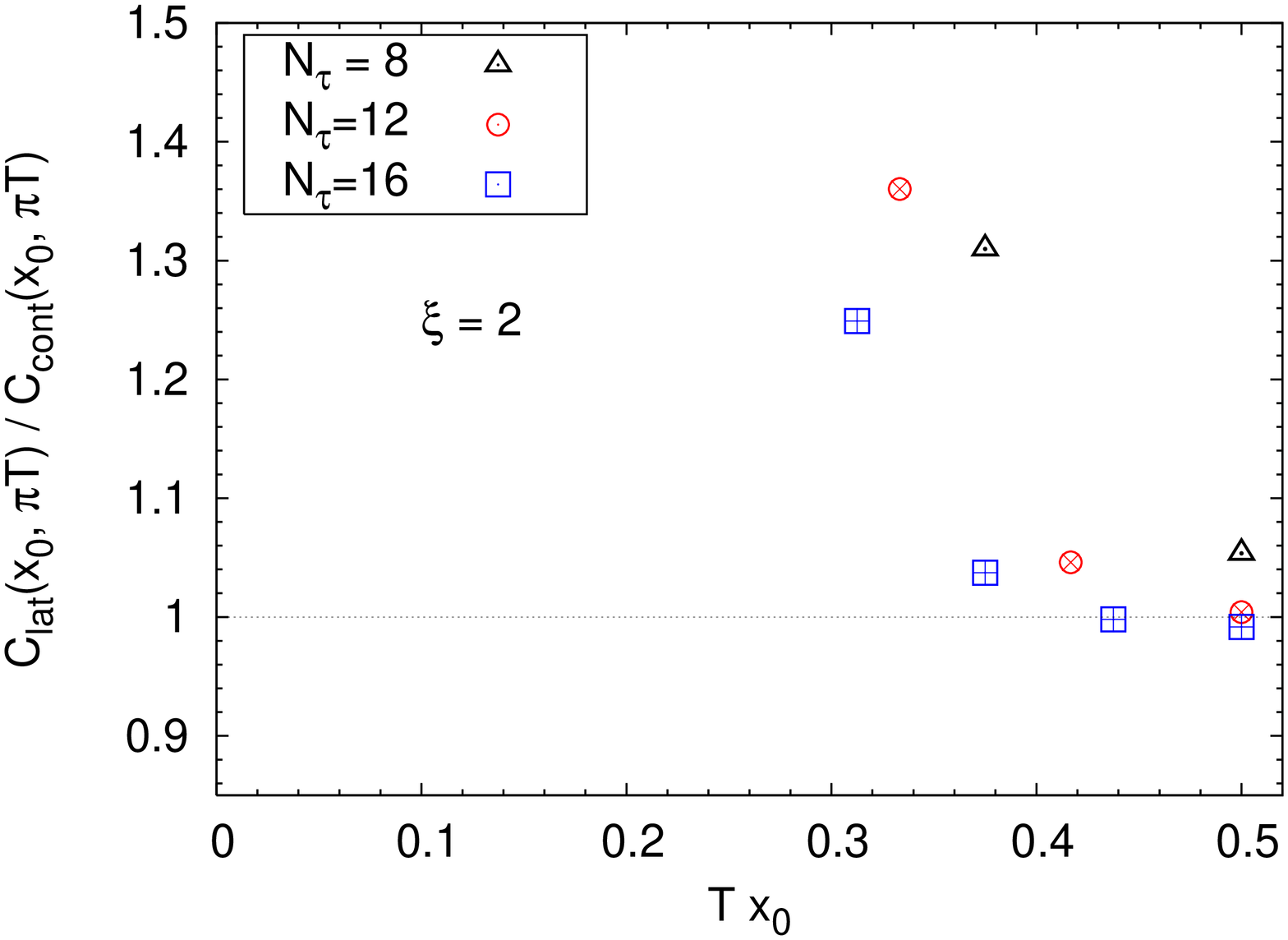,angle=-90,width=13cm}}
\centerline{\psfig{file=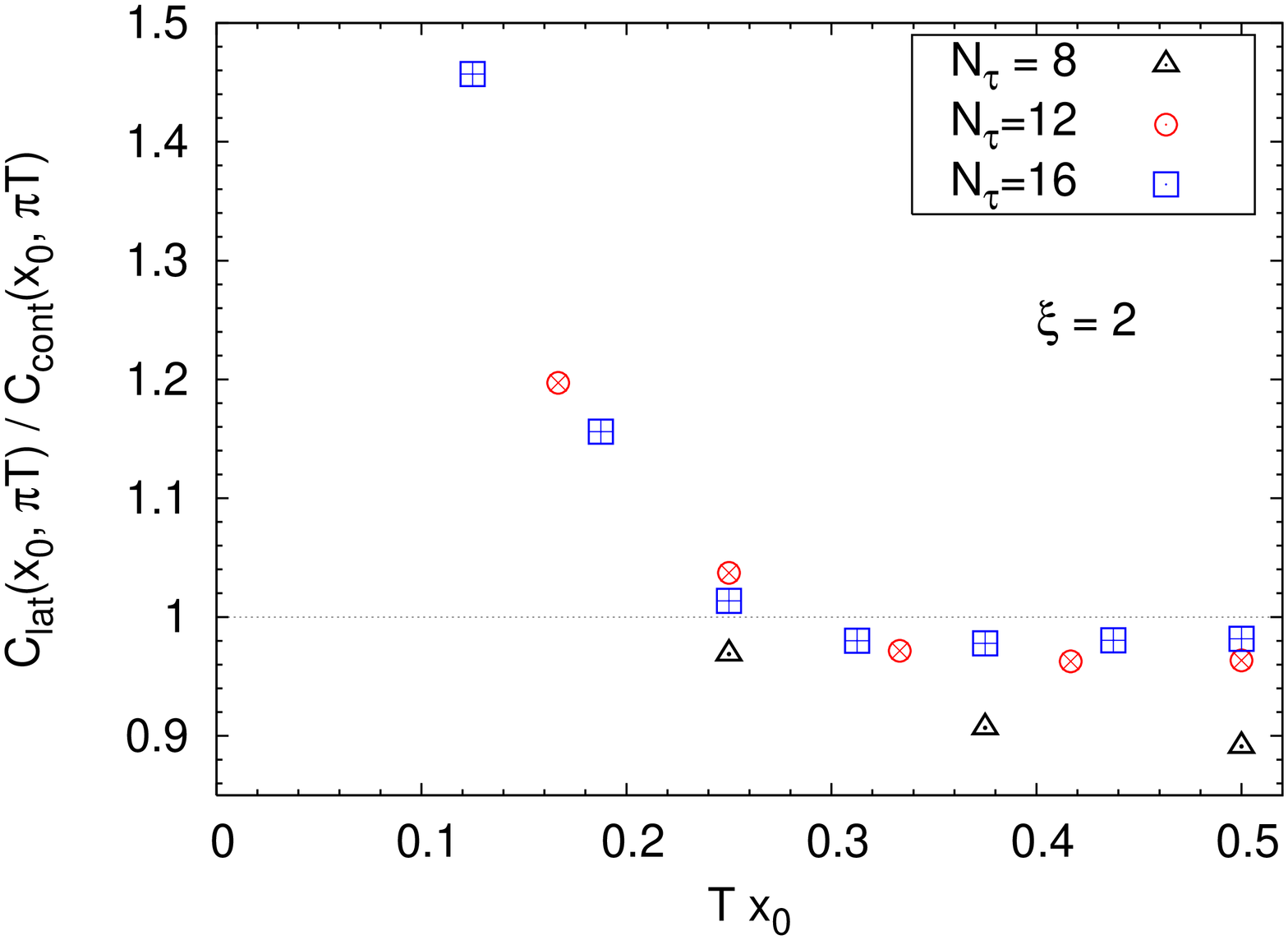,angle=-90,width=13cm}}
\caption{\small  The ratio of lattice to continuum treelevel correlators for 
$T_{00}$ (top) and $T_{03}$ (bottom). 
Clover discretization on the anisotropic lattice, for ${\bf p}=(0,0,\pi T)$.
}
\la{fig:momentum}
}

\section{Treelevel improvement\la{sec:tlimp}}
Here we describe how the results obtained in this paper can be used to remove the dominant
part of the cutoff effects on the correlators. In the case of the clover discretization,
we simply divide by the treelevel lattice result, and multiply by the continuum result:
\be
C_{\rm lat}(x_0,{\bf p})  \to  C_{\rm lat}(x_0,{\bf p})  
\,\cdot\, \frac{C^{\rm t.l.}_{\rm cont}(x_0,{\bf p}) }{C^{\rm t.l.}_{\rm lat}(x_0,{\bf p})  }
\ee
This technique is not new, see for instance \cite{Gimenez:2004me}.

For the plaquette discretization,
the three electric-electric, magnetic-magnetic and electric-magnetic
contributions to $C(x_0)$ are computed separately.
For each of tem, we apply the following technique~\cite{Meyer:2007ic} 
to remove the tree-level discretization errors,
which is adapted from static-potential studies~\cite{Sommer:1993ce}.
Firstly, $\bar x_0$ is defined by the treelevel correlators such that 
$C_{\rm cont}^{\rm t.l.}(\bar x_0)=C_{\rm lat}^{\rm t.l.}(x_0)$.
The improved non-perturbative correlator $\overline C$ is defined at a discrete set of points through
$\overline C(\bar x_0) = C( x_0)$, and then augmented to a continuous function
$\overline C= \alpha + \gamma C_{\rm cont}^{\rm t.l.}$. The parameters $\alpha$ and $\gamma$
are fixed by the condition
$\overline C(\bar x_0^{(i)}) = \alpha + \gamma C_{\rm cont}^{\rm t.l.}(\bar x_0^{(i)}) $, $i=1,2$,
where $\bar x_0^{(1)}$ and $\bar x_0^{(2)}$ correspond to two adjacent measurements. 
The correlator is thus defined for all
Euclidean times between $\bar x_0^{(1)}$ and $\bar x_0^{(2)}$.
The linear combination of the  electric-electric, magnetic-magnetic and electric-magnetic
contributions is finally obtained at a common value of Euclidan time.
Derivatives can then also be obtained from
 $\gamma \frac{d^n}{dx_0^n}C_{\rm cont}^{\rm t.l.}(x_0)$.

It is clear from this discussion that it is simpler to have a site-centered
definition of the energy-momentum tensor, and this is the choice made for our large-scale
calculation~\cite{Meyer:2008sn}.

\section{Non-perturbative study of cutoff effects}
In this section we test how effective treelevel improvement is, 
by applying it to non-perturbative data from Monte-Carlo simulations.

\subsection{Isotropic lattice, plaquette discretization}
We start with data obtained on isotropic lattices with the plaquette discretization.
In the scalar channel, the $C_{\theta\theta}$ correlator was obtained at several lattice spacings 
in~\cite{Meyer:2007ic}.
Figure \ref{fig:Sconti} displays the lattice spacing dependence of the $Tx_0=\frac{1}{2}$
treelevel-improved \emph{tensor} correlator as a function of $(aT)^2$ for two fixed temperatures.
The aspect ratio of the lattice is $LT=5$ at $1.65T_c$ and $LT=2.5$ for $1.24T_c$.
The residual discretization errors appear to be small, in fact consistent
with zero  for $\Nt\geq8$. However, a residual $10\%$ cutoff effect 
cannot be excluded at $1.65T_c$, because the statistical errors are increasing with $\Nt$.

\subsection{Anisotropic lattice ($\xi=2$), clover discretization}
Figure \ref{fig:Sconti} displays the treelevel-improved correlators of the 
total energy and momentum on a $16\times48^3$ lattice. As discussed in section 3.1, these 
correlators are $x_0$-independent in the continuum limit.
In the case of the momentum $T_{0k}$, the correlator is flat within statistical errors
down to $Tx_0=\frac{1}{4}$ or $\frac{1}{5}$. Since this corresponds to 
$x_0/\as=1.5$ and 2 respectively, we regard this as an excellent outcome.
By contrast, the unimproved correlator starts to rise for $x_0/\at$ larger by one unit.

In the case of the energy, the treelevel improvement imposes a much larger
correction to the data, and, not surprisingly, the expected flatness of the 
correlator is much less well realized. The treelevel improvement undercorrects the correlator
for $Tx_0\geq \frac{1}{4}$, and overcorrects it at short distances.
We have to conclude that in this channel, only the largest $x_0/\at$ points are
usable for a continuum extrapolation, even after treelevel improvement.
In particular, this strongly restricts  the $x_0$-information for finite-${\bf p}$ 
correlators of the energy density. The latter are particularly interesting~\cite{Meyer:2008gt},
because they contain information on the damping of sound waves in the plasma.

\FIGURE[t]{
\centerline{\psfig{file=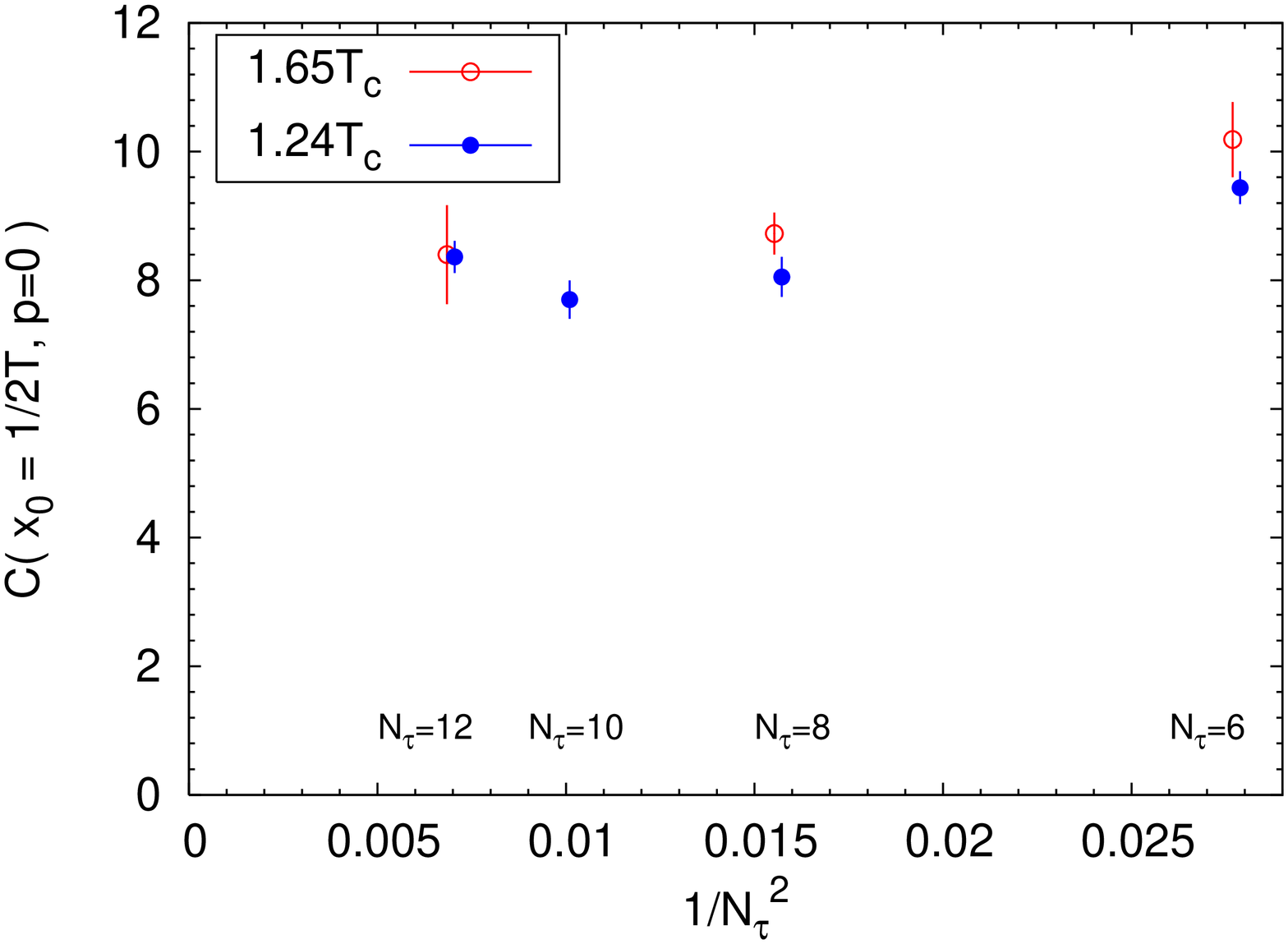,angle=-90,width=13cm}}
\centerline{\psfig{file=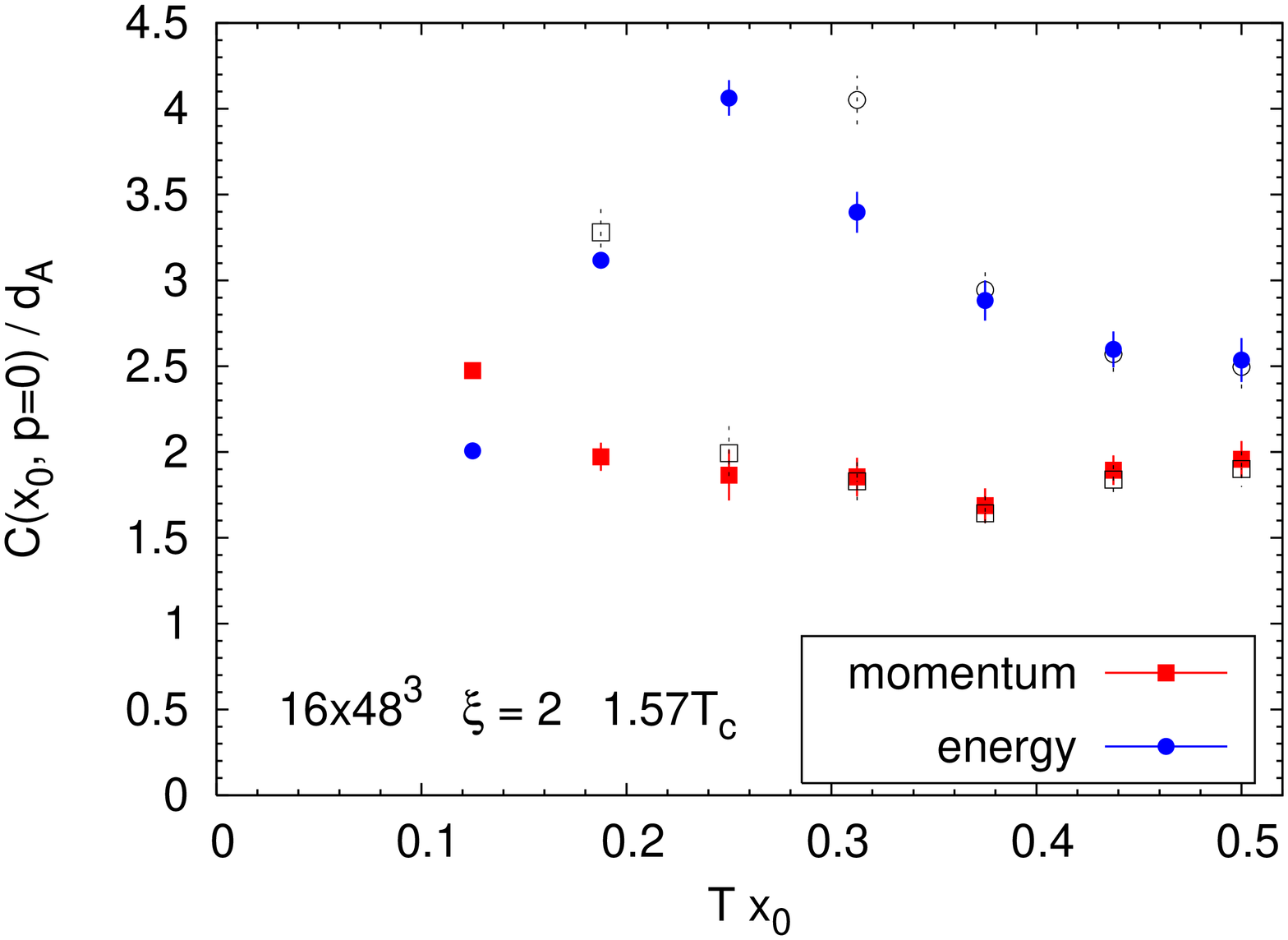,angle=-90,width=13cm}}
\caption{{\bf Top}:  The cutoff effects on the 
$\half(T_{11}-T_{22})$ 
treelevel-improved two-point function from Monte-Carlo 
simulations, for two temperatures, $1.65$ and $1.24T_c$.
Here the plaquette discretization is used  on the isotropic lattice.
{\bf Bottom}: The treelevel-improved 
correlators $C_{00,00}$ and $\sum_k C_{0k,0k}$ at ${\bf p}=0$
calculated with the clover discretization on the anisotropic lattice at $1.57T_c$.
These correlators are flat and equal to $c_v$ and $3s$ resp. in the continuum limit.
The unimproved correlators are displayed as open symbols.
}
\la{fig:Sconti}
}

\section{Concluding remarks}
We have studied the discretization errors of the energy-momentum tensor correlators
in lattice gauge theory. We summarize the lessons learnt.
\begin{enumerate}
\item using a site-centered discretization of the EMT  simplifies the calculation 
of correlators and their treelevel improvement. 
Therefore in the following we discuss the `clover' discretization.
\item the momentum density correlator has small cutoff effects at treelevel, 
and correspondingly the treelevel-improved non-perturbative correlator
has small cutoff effects down to very small separations.
\item the energy density correlator is much more problematic with the chosen 
discretization. The treelevel energy correlator is far from being flat for realistic 
values of $\Nt$, and correspondingly after treelevel improvement  cutoff effects
as large as $50\%$ remain. This implies that only the largest values of $x_0$ can be 
used in a continuum extrapolation.
\item the correlators of the spatial components of $T_{\mu\nu}$ have moderate
discretization errors and treelevel improvement works well.
\item in all analyzed channels,
the anisotropic lattice helps reduce the cutoff effects at a lower cost
than decreasing the lattice spacing on the isotropic lattice.
\end{enumerate}

The present study suggests that the design of a lattice energy density operator 
which leads to small cutoff effects in its two-point functions 
would be very valuable. Such a discretization would also have an impact
in other areas of QCD, for instance in hadron structure calculations, 
where the glue energy density operator determines 
the glue momentum fraction~\cite{Gockeler:1996zg,Meyer:2007tm}.
It would be interesting to investigate the cutoff effects associated with
the HYP-smeared discretizations introduced in~\cite{Meyer:2007tm}.

A further important issue on the anisotropic lattice
is the proliferation of normalization 
factors that have to be applied to different components of the energy-momentum 
tensor. Their determination will be the subject of a separate publication.

\acknowledgments{
Lattice computations for this work were partly carried out on facilities of
the USQCD Collaboration, which are funded by the Office of Science of
the U.S. Department of Energy, and partly on the BlueGeneL at MIT. 
This work was supported in part by funds provided by the U.S. Department of Energy 
under cooperative research agreement DE-FG02-94ER40818.
}

%%%%%%%%%%%%%%%%%%%%%%%%%%%%%%%%%%%%%%%%%%%%%%%%%%%%%%%%%%%%%%%%%%%%%%%%%%%
\bibliographystyle{JHEP}
\bibliography{/afs/lns.mit.edu/user/meyerh/BIBLIO/viscobib}
%%%%%%%%%%%%%%%%%%%%%%%%%%%%%%%%%%%%%%%%%%%%%%%%%%%%%%%%%%%%%%%%%%%%%%%%%%%%

\end{document}